\documentclass[10pt,onecolumn,oneside,a4paper]{article}
\usepackage[english]{babel}  
\usepackage{amsmath,amssymb,amsfonts,amsthm,dsfont}
\usepackage{authblk}
\usepackage{stackengine}
\usepackage{verbatim}
\usepackage{longtable}
\usepackage{algorithm,algorithmicx,algpseudocode,float}
\usepackage{hyperref}
\usepackage{graphicx}
\usepackage{xcolor}
\usepackage{url}
\PassOptionsToPackage{hyphens}{url}\usepackage{hyperref}
\makeatletter
\g@addto@macro{\UrlBreaks}{\UrlOrds}
\makeatother

\usepackage[font=small,skip=6pt]{caption}

\makeatletter
\newenvironment{breakablealgorithm}
  {
  \begin{center}
     \refstepcounter{algorithm}
     \hrule height.8pt depth0pt \kern2pt
     \renewcommand{\caption}[2][\relax]{
      {\raggedright\textbf{\ALG@name~\thealgorithm} ##2\par}%
      \ifx\relax##1\relax 
         \addcontentsline{loa}{algorithm}{\protect\numberline{\thealgorithm}##2}%
      \else 
         \addcontentsline{loa}{algorithm}{\protect\numberline{\thealgorithm}##1}%
      \fi
      \kern2pt\hrule\kern2pt
     }
  }{
     \kern2pt\hrule\relax
  \end{center}
  }
\makeatother

\floatname{algorithm}{Procedure}

\renewcommand{\algorithmicreturn}[1]{\bgroup\\  ~#1\egroup}
\renewcommand{\algorithmiccomment}[1]{\bgroup\hfill//~#1\egroup}

\newcommand{\graytext}[1]{\color{black}{}\color{gray}{#1}\color{black}{}}

\begin{document}


\newpage

\title{Neural Network Training With Homomorphic Encryption}

\author[1]{Kentaro Mihara}
\author[1,2]{Ryohei Yamaguchi}
\author[1,3]{Miguel Mitsuishi}
\author[1,3]{Yusuke Maruyama}

\affil[1]{\small{EAGLYS Inc, Research and Development, Tokyo, Japan}}
\affil[2]{\small{Tokyo University of Agriculture and Technology, Department of Mechanical Systems Engineering, Tokyo, Japan}}
\affil[3]{\small{Waseda University, Faculty of Science and Engineering,  Tokyo, Japan}}

\maketitle

\begin{abstract}
   We introduce a novel method and implementation architecture to train neural networks which preserves the confidentiality of both the model and the data. Our method relies on homomorphic capability of lattice based encryption scheme. Our procedure is optimized for operations on packed ciphertexts in order to achieve efficient updates of the model parameters. Our method achieves a significant reduction of computations due to our way to perform multiplications and rotations on packed ciphertexts from a feedforward network to a back-propagation network. To verify the accuracy of the training model as well as the implementation feasibility, we tested our method on the Iris data set by  using the CKKS scheme with Microsoft SEAL as a back end. Although our test implementation is for simple neural network training, we believe our basic implementation block can help the further applications for more complex neural network based use cases.

\end{abstract}

\section{Introduction}\label{sec:Intro}

Deep learning has become successful mostly because of hardware improvements and cloud computing. Besides the huge quantity of data that can be processed, it remains many issues regarding safety such as privacy and anonymity related issues. Cloud computing in its essence leaks information since private information leaves physically or systematically secure environment such as server reside in the company or institution. In the realm of deep learning, two important kinds of information can be leaked: the parameters that define the model and the data used to to train the model. Privacy-preserving machine learning have been paid attention over the last decade in order to prevent such leakage.

Privacy-preserving machine learning methods can be categorized into two main classes: secure multiparty computation, and homomorphic encryption. A secure multiparty approach creates a generic training protocol that allows several parties to compute a function on their joint inputs without sharing secrets. Mohassel and Zhang~\cite{MhoZha_2017} presented privacy-preserving neural network training protocol based on a system with two untrusted servers. Although they show the feasibility of their protocol, the latter is very expensive both computationally and in communication. A homomorphic approach does computations over encrypted data as well as the encrypted model, or just over encrypted data alone if there is no need to conceal the model parameters. Homomorphic operations require only one server in general to train the model while a multiparty computation based approach would require the simultaneous interactions of all the parties involved. Research on machine learning inferential applications using homomorphic encryption can be found for instance in~\cite{badawi2020alexnet, BouMinMinPai_2018, CHET_2019, JiaKimLauSon_2018,  JuvVaiCha_2018, LiuJuuLuAso_2017, MhoZha_2017, RouRiaKou_2018, SakLu_2017}. Applications for deep machine learning are however not so frequent~\cite{KimSonKimLeeChe_2018}.

\section{Preliminaries}\label{sec:Sec2}

In this section, we shall recall the CKKS~\cite{CKKS_2017} scheme upon which our new application is based. Since our implementation uses the CKKS scheme backed with the Microsoft SEAL library~\cite{SEAL}, we also briefly summarize the latter's functionalities. In addition, we review packing methods for that are crucial for machine learning applications that use homomorphic encryption. We review quickly the necessary material for training neural networks that motivated our work.

A CKKS scheme has its security based on the hardness of learning with errors (LWE) problem over a polynomial factor ring ~\cite{RLWE_2010}. It has a different encoding and batching scheme than the BFV scheme~\cite{FV_2012}. One of the major difference is that CKKS scheme allows us to encode real number into polynomial slot through canonical embedding whereas BFV scheme allows only integer for encoding, CKKS scheme is preferable for our implementation. It supports leveled homomorphic encryption with chained modulus and relinearization as with the BFV scheme. The main functionalities are: (1) KeyGen, (2) Encoding, (3) Decoding, (4) Encryption, (5) Decryption, (6) Addition, and (7) Multiplication and (8) Rotation.

Packing and rotation techniques are central to compute over ciphertexts because they can be used to speed up linear transformations significantly. Packing is used to encrypt a list of elements to one ciphertext instead of encrypting each element into distinct ciphertexts. This is done through encoding the plaintext information into the subfield of the factor ring~\cite{SmaVer_2014}. Rotation is the Galois mapping of each element from the subfield, which in practice, is used to add elements from one subfield to another. In this way, linear transformation can be written as a sequence of sums, multiplications and rotations over packed ciphertext. Multiplications and rotations are computationally heavy operations, hence, the design of efficient ciphertext packing methods are critical. We will study the effect of different packing methods on the computation load.


We focus now on how to perform matrix-vector multiplications efficiently using two packing algorithms: (1) a row vector packing method, and (2) a diagonal packing method. These two packing methods are useful to perform linear transformations over encrypted vector. First, consider the multiplication of a matrix $W$ with a vector $x$ whose dimensions are $M\times N$ and $N$, respectively. Following procedure is summarized in algorithm~\ref{m-v-row}. Here, we denote $\texttt{rot}(x, i)$ for packed vector $x$ rotated to right direction by $i$ slots. Similarly $\texttt{rot}(x, -i)$ is for packed vector $x$ rotated to left direction by $i$ slots. For instance, $\texttt{rot}([1,2,3,4,5], 1) = [5,1,2,3,4]$,  $\texttt{rot}([1,2,3,4,5], -1) = [2,3,4,5,1]$. As shown on figure~\ref{fig:prelim1} below, row vector packing method packs each row of $W$ respectively which results in $M$ packed ciphertexts for $W$ and one for $x$. After packing, multiplication is processed by multiplying each of the $M$ packed ciphertexts with the packed ciphertext $x$.
\begin{figure}[h]
\begin{centering}
	\includegraphics[width=0.75\textwidth]{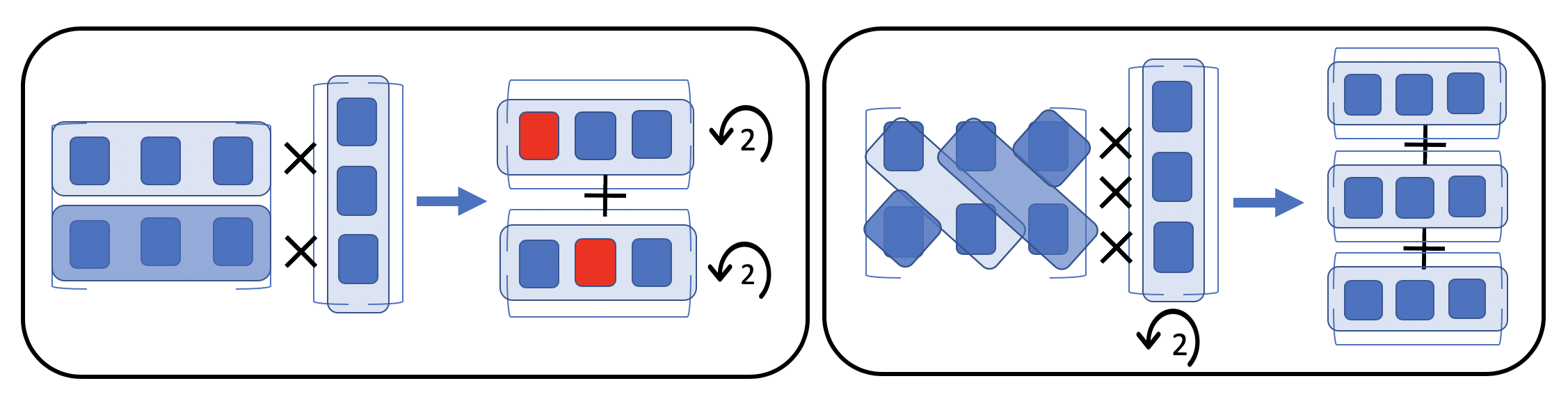}
	\caption{Matrix-vector multiplication with row  packing method (left) and diagonal packing method (right). Rotation operation is depicted as curved arrow with operation number written inside.}\label{fig:prelim1}
\end{centering}
\end{figure}

In order to obtain the final result, each intermediate results are rotated and summed by itself so that the slot contains the inner product value of row-packed vector and $x$. Once each inner products are obtained, multiplication with the vector with only one element 1 and the others 0 (we note this vector as unit vector) is applied to remove unnecessary elements. At the end, results are summed up to yield the output of linear transformation. In this case, this entire process consumes $2M$ multiplications and $MN$ rotations.

We describe now the diagonal packing method from \texttt{gazelle}~\cite{JuvVaiCha_2018}. Given a fixed matrix $W$ and and fixed input $x$, packing is done along the diagonal direction instead of row direction as shown on figure~\ref{fig:prelim1} (right). Input, on the other hand, packed into one ciphertext as before. In this case, we conduct iteration which multiplies each of packed diagonal component of the weight $W$ with rotated packed input vector. Packed $x$ is rotated by one at each iteration step. After this iteration, a summation of those intermediate results yields to linear transformation. Major difference from row packing method is that there are no need to multiply ciphertext with the unit vector. In addition, since rotation is only applied to the input vector, the number of rotation can be fewer. Hence, the total computation cost is composed of $N$ multiplications and $N$ rotations. Procedure is summarized in algorithm ~\ref{m-v-diagonal}. Although the number of multiplication depends on the shape of given matrix and of the input vector, diagonal method is more efficient in most of the situations.

 \begin{breakablealgorithm}\label{m-v-row}
 \caption{Algorithm of matrix-vector multiplication with row packing}
 \begin{algorithmic}[1]
 \raggedright
 \Require $\textbf{W}=\{\textbf{w}_i,0\le i\le M-1 \},\textbf{x}, \textbf{U}=\{\textbf{u}_i,0\le i\le M-1 \}$
 \Ensure  \textbf{P}
  \For{$i = 0$ to $M-1$}
  \State $\textbf{y1} = \sum_{s=0}^{s=N-1} \texttt{rot}(\textbf{w}_i \times \textbf{x}, -s)$
  \State $\textbf{y2} = \texttt{rot}(\textbf{y1}, i) \times \textbf{u}_i $

  \EndFor
 \State $\textbf{P} = \sum_{s=0}^{s=M-1} \textbf{y2}_s$
 \State{\textbf{RETURN} $\textbf{P}$} 
 \end{algorithmic} 
 \end{breakablealgorithm}
 
 \begin{breakablealgorithm}\label{m-v-diagonal}
 \caption{Algorithm of matrix-vector multiplication with diagonal packing}
 \begin{algorithmic}[1]
 \raggedright
 \Require $\textbf{W}=\{\textbf{w}_i,0\le i\le N-1 \}$
 \Ensure  \textbf{P}
  \State $\textbf{P} = \sum_{i=0}^{i=N-1} \textbf{w}_i \times \texttt{rot}(\textbf{x}, -i)$
 \State{\textbf{RETURN} $\textbf{P}$} 
 \end{algorithmic} 
 \end{breakablealgorithm}

\begin{figure}[h]
\begin{centering}
	\includegraphics[width=0.75\textwidth]{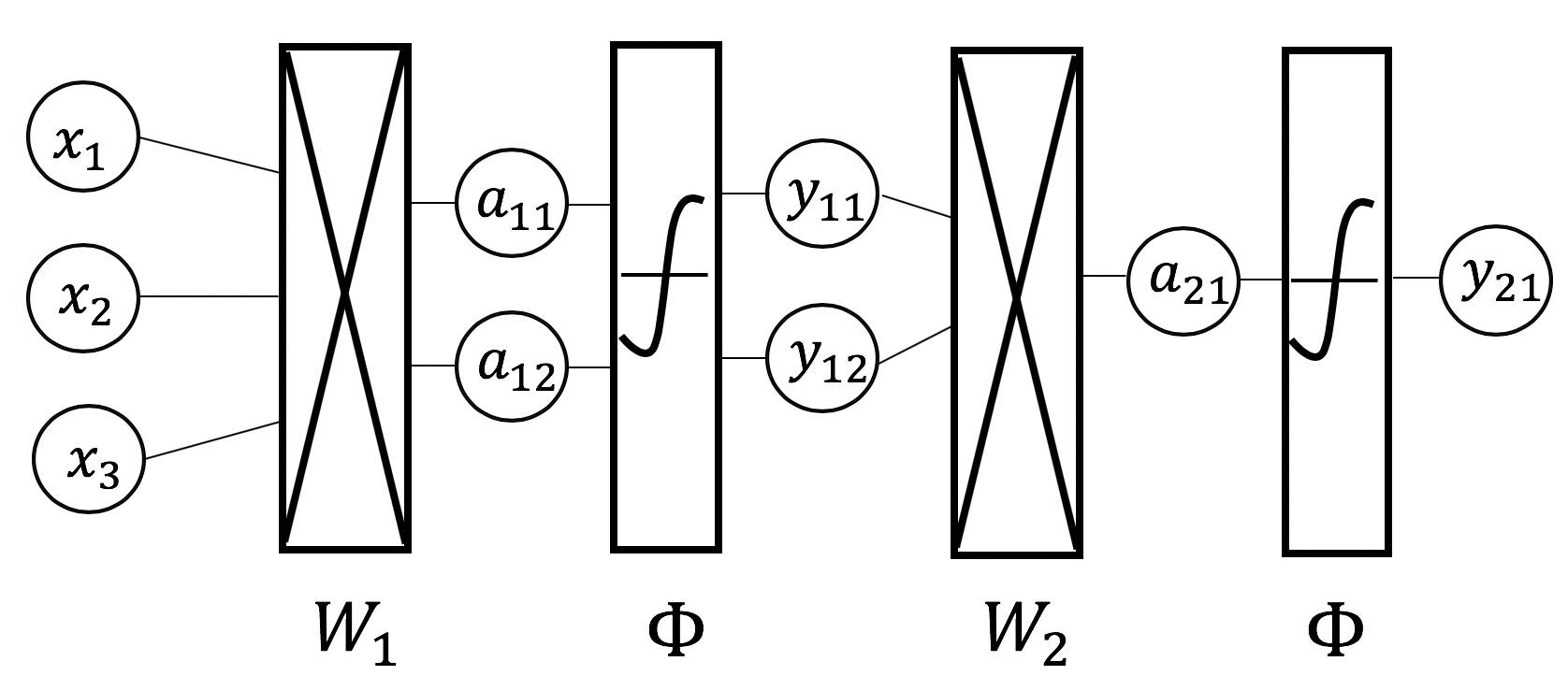}
	\caption{Example of simple neural network architecture with linear transformation dense layer and non-linear activation layer. In this figure, input vector with dimension of $3$ is affine transformed by weight matrix $\mathbf{W}_{1}$ and path the activation layer $\phi$. It repeats this process to the end of the network.}\label{fig:prelim2}
\end{centering}
\end{figure}

We briefly review neural network training procedures. For simplicity, we show a simple neural network model which consists of three layers: the input layer, the middle hidden layer and the output layer. A nonlinear layer called activation layer is placed after each linear transformation. The generic structure of a neural network is as on figure~\ref{fig:prelim2}.

Neural network training is decomposed into two main parts: feedforward and backpropagation. Feedforward is made by the composition of a linear transformation layer followed by a non-linear transformation layer, called activation layer. We initialize the entries of the weight matrix and the biases using a gaussian distribution. The dataset which consists of an input vector and a label that is fed to the first layer then linearly transformed using the weights and the biases. An output is transformed non-linearly using the activation layer. The output of the activation layer becomes the input for the second layer and this continues until we reach the last layer. The output from the last layer becomes the prediction result of that specific input vector. In order to improve the prediction, a the loss between the inferred results and the actual labels is calculated and passed to the backpropagation. Backpropagation is generally the reverse process of feedforward. For the backpropagation process, updates of the weights are calculated to optimize the model towards a better prediction. On figure~\ref{fig:prelim3} below, the input goes through two layers by feedforward and the loss is evaluated by taking the mean square error between the inferred result and the true label. Once the loss is calculated, backpropagation allows to compute partial derivatives of the loss with respect to parameters. For the update of the weight parameter, this partial derivative is used to minimize the loss function over the training. Along this backpropagation step, the transpose of packed encrypted weight matrix is required figure~\ref{fig:prelim3}

\begin{figure}[h]
\begin{centering}
	\includegraphics[width=0.75\textwidth]{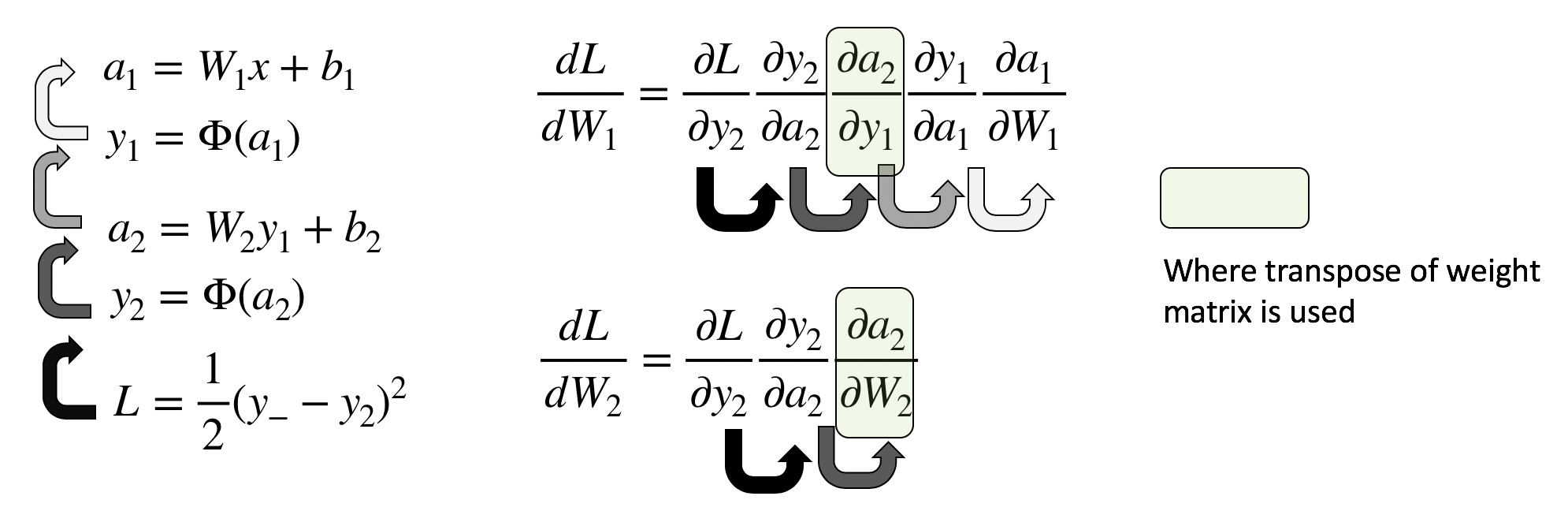}
	\caption{Backpropagation with transpose weight matrix}\label{fig:prelim3}
\end{centering}
\end{figure}

Optimizations of the following primitive operations are thus crucial to perfect efficiently neural network training with homomorphic encryption: (1) matrix-vector operations to perform the feedforward phase, (2) computation of the transposition  weight matrix for the backpropagation, (3) matrix-vector operations within the backpropagation step, and (4) method to computer the activation function. For (1) and (3), as proposed in \texttt{gazelle}~\cite{JuvVaiCha_2018} implementation, we pack along the diagonal. Square activation is used for (4) and also as a homomorphic friendly activation function as in \texttt{cryptonets}~\cite{CryptoNets_2016, KimSonKimLeeChe_2018}. We discuss (2) in the next section since the efficient transposition of packed matrix is not discussed in the previous literature.

\section{Results and related work}\label{sec:Sec3}

We mentioned previously that the optimization of the computation of the linear transformation is important in order to keep the runtime and the memory size small enough to be practical. Optimization efforts for homomorphic operations on machine learning algorithms such as convolutional neural network are done by~\cite{JiaKimLauSon_2018, JuvVaiCha_2018, chao2019carenets} which also make use of packing method for the inputs. However, the aforementioned articles focus on the inference part rather than training the model. Recently,~\cite{KimSonKimLeeChe_2018} proposed an attempt to train a logistic regression model using packed information to allow end-to-end training. Nevertheless, to the best of our knowledge, only a few implementations of neural network training have been done. \emph{Our contribution} is as follow:
\begin{enumerate}
\item A novel packing method for the matrix of weights to achieve efficient overall training runtime. As a result, our design keeps the total number of heavy homomorphic operations significantly lower than other methods. In addition, the depth of circuit for multiplication is kept small which yields to faster and lighter implementation.
\item A detailed implementation which was not done extensively in the industry before.
\item A tested implementation of our method on well-known iris dataset using Microsoft SEAL.
\end{enumerate}

\subsection{Our matrix transposition method}\label{sec:Sec4}

In this section, we explain our algorithm which transpose the packed matrix of weights that occurs during the transition from feedforward to backpropagation. Diagonal packing method is adopted throughout the training procedure due to the following reasons:
\begin{enumerate}
    \item\label{reason1} Number of rotations is intrinsically smaller.
    \item\label{reason2} Difference between multiplication of row and diagonal packing is compensated by the training procedure.
    \item\label{reason3} No multiplication is needed for creating transpose matrix, therefore, no extra circuit depth is necessary for taking transpose.
\end{enumerate}
For (\ref{reason1}), as is mentioned previously, the number of rotation operations upon linear transformation can be reduced with diagonal packing. For (\ref{reason2}), although it is mentioned that the number of multiplication depends on both the input and output dimensions for the linear transformation, these differences are compensated because the training process has forward and backward propagations. The most significant point is (\ref{reason3}) because, with our method, diagonal packed weight does not require any multiplications but only rotations to transpose the packed matrix. As a result, our configuration can have smaller modulus chain for multiplication circuit, which significantly benefit us for computationally lighter design.

\begin{figure}[h]
\begin{centering}
	\includegraphics[width=0.7\textwidth]{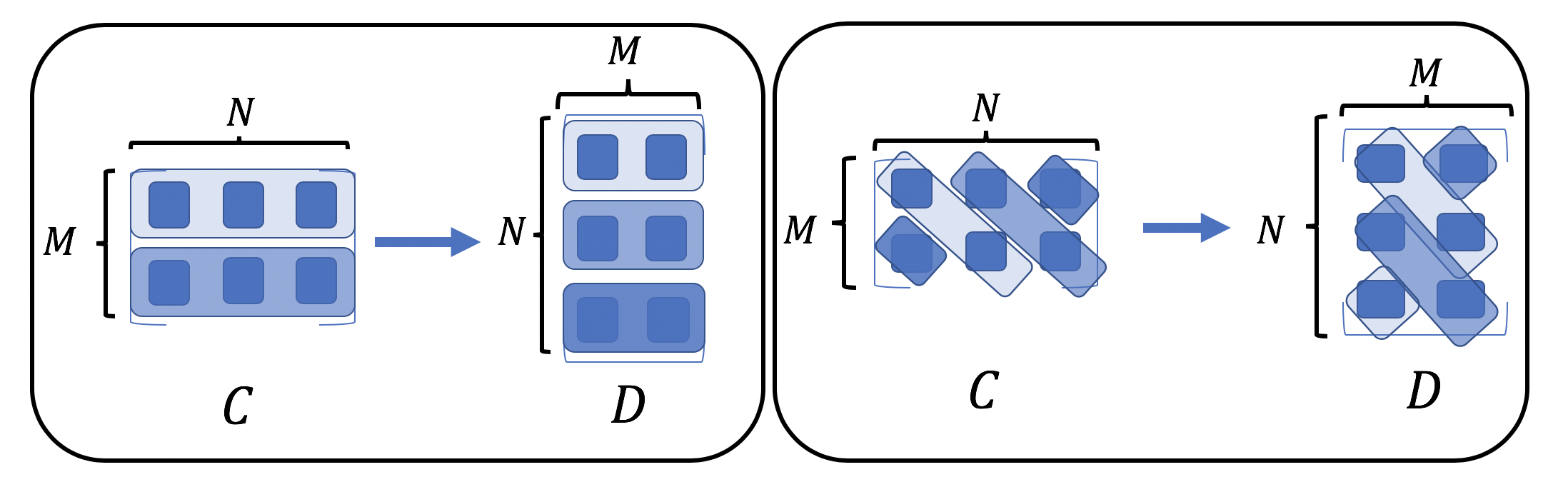}
	\caption{\label{fig:ourtransitionmethod1}Transpose of packed matrix with row packing (left) and diagonal packing (right). In order to make transpose matrix, diagonal method has huge advantage because packed ciphertext can be reused.}
\end{centering}
\end{figure}

For instance, let the input dimension of a dense weight matrix be $N$ and its output dimension be $M$, that is the weight matrix has size $N\times M$. We denote the original packed weight matrix by $\mathbf{C}$ and the packed transpose weight matrix by $\mathbf{D}$. Hence $\mathbf{C}$ is consists of $M$ ciphertexts, each of them are packed row of original matrix. Also $\mathbf{D}$ consists of $N$ ciphertexts, each of them are original packed components of the transpose weight matrix as shown in figure ~\ref{fig:ourtransitionmethod1}. If we adopt row packing for packing algorithm, making components of $\mathbf{D}$ from $\mathbf{C}$ takes (1) multiplication by unit vector, (2) rotation, and (3) summation. This procedure is shown in algorithm \ref{transpose-row}. We observe that the fact that components of $\mathbf{C}$ and $\mathbf{D}$ have few structural similarities made this procedure tedious. In the example, $N\times M$ multiplications and $N\times M$ rotations are needed in total to generate $\mathbf{D}$ out of $\mathbf{C}$.

 \begin{breakablealgorithm}\label{transpose-row}
 \caption{Algorithm of taking transpose with row packing}
 \begin{algorithmic}[1]
 \raggedright
 \Require $\textbf{C}=\{\textbf{c}_i,0\le i\le M-1 \}$
 \Ensure  $\textbf{D}=\{\textbf{d}_i,0\le i\le N-1 \}$
  \For{$i = 0$ to $M-1$}
     \For{$j = 0$ to $N-1$}
       \State $\textbf{y1}[i][j] = \texttt{rot}(\textbf{c}_i \times \textbf{u}_j, -j+i)$
    \EndFor
  \EndFor
  \State $\textbf{d}_k = \sum_{i=0}^{i=M-1} \textbf{y1}[i][k]$
 \State{\textbf{RETURN} $\textbf{D}$} 
 \end{algorithmic} 
 \end{breakablealgorithm}

On the other hand, with diagonal method, packed components $\mathbf{C}$ have good structural similarities with the packed components of transposed matrix $\mathbf{D}$. Therefore, creating $\mathbf{D}$ out of $\mathbf{C}$ is a lot easier: (i) no multiplication by unit vector, (ii) less rotation needed from row packing case. As in example, 0 multiplication and $q(M+1)$ rotation needed where $N$ and $M$ has a relationship as $N = qM + r$.

\begin{figure}[h]
\begin{centering}
	\includegraphics[width=0.75\textwidth]{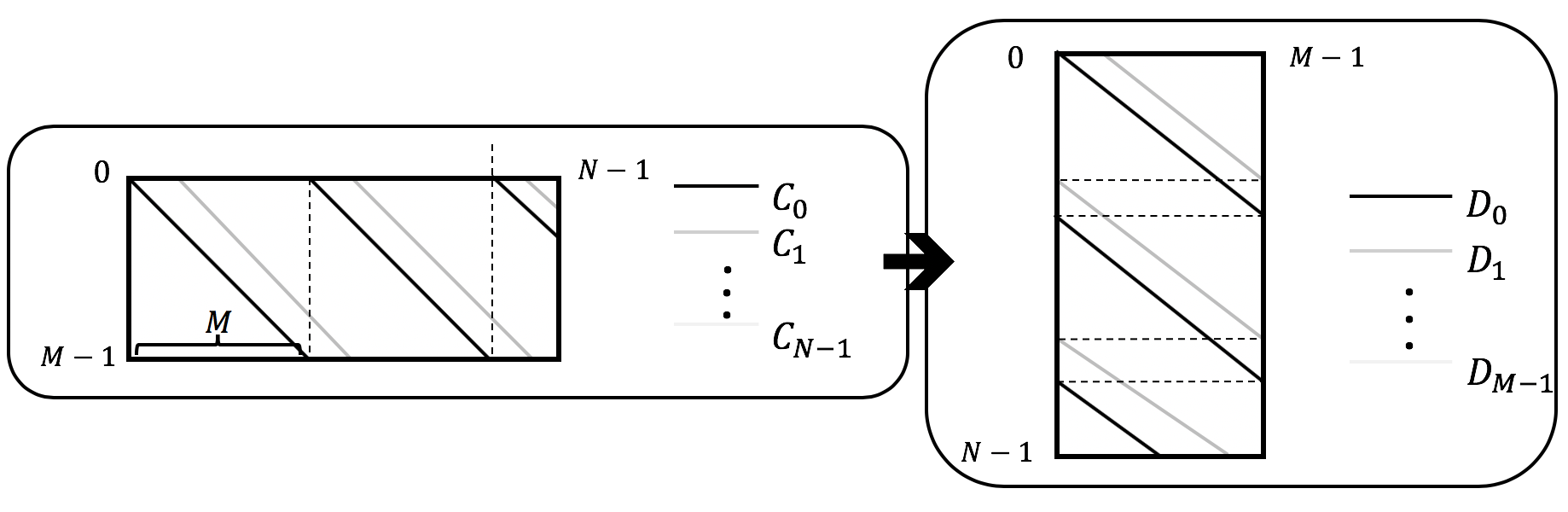}
	\caption{\label{fig:ourtransitionmethod2}Generalized algorithm for making transpose of packed matrix with diagonal method.}
\end{centering}
\end{figure}

Not only this algorithm significantly reduces the multiplication cost in total computation, it also keeps the circuit depth lower for training since no multiplication consumed. Therefore, it allows us to execute the training with relatively lighter encryption configuration, which helps reduces computational overhead significantly.
We generalized this algorithm for generation of packed transposed matrix out of packed matrix with only rotation and summation on figure~\ref{fig:ourtransitionmethod2}.
If we denote $x$ rotated $A$ to be the $x$ rotated in right direction of originally packed ciphertext $A$,  on the right side of figure~\ref{fig:ourtransitionmethod2}, $D_0$ can be expressed with summation of $C_0$, $\texttt{rot}(C_M, M)$ and $\texttt{rot}(C_{2M}, 2M)$. $D_1$, however, is little bit more tricky, can be expressed with the summation of $\texttt{rot}(C_{N-1}, -1)$,  $\texttt{rot}(C_{M-1}, M-1)$ and  $\texttt{rot}(C_{2M-1}, 2M-1)$. Following these, all the transposed packed matrix $D$ can be expressed by rotation and summation of $C$ as follows. The generalized equation is summarized below.

\begin{align*}
e_i &=\left\{
\begin{array}{ll}
q& \text{if $0\leq i \leq M-r$} \\
q+1 &\text{if $M-r < i \leq M$}
\end{array}\right.\\
D_i & = \sum_{s=0}^{s=e_i}{\texttt{rot}\big(C_{sM-i}, sM-i\big)}\quad\text{where $N = qM + r$.}
\end{align*}

\begin{figure}[h]
\begin{centering}
\includegraphics[width=0.6\textwidth]{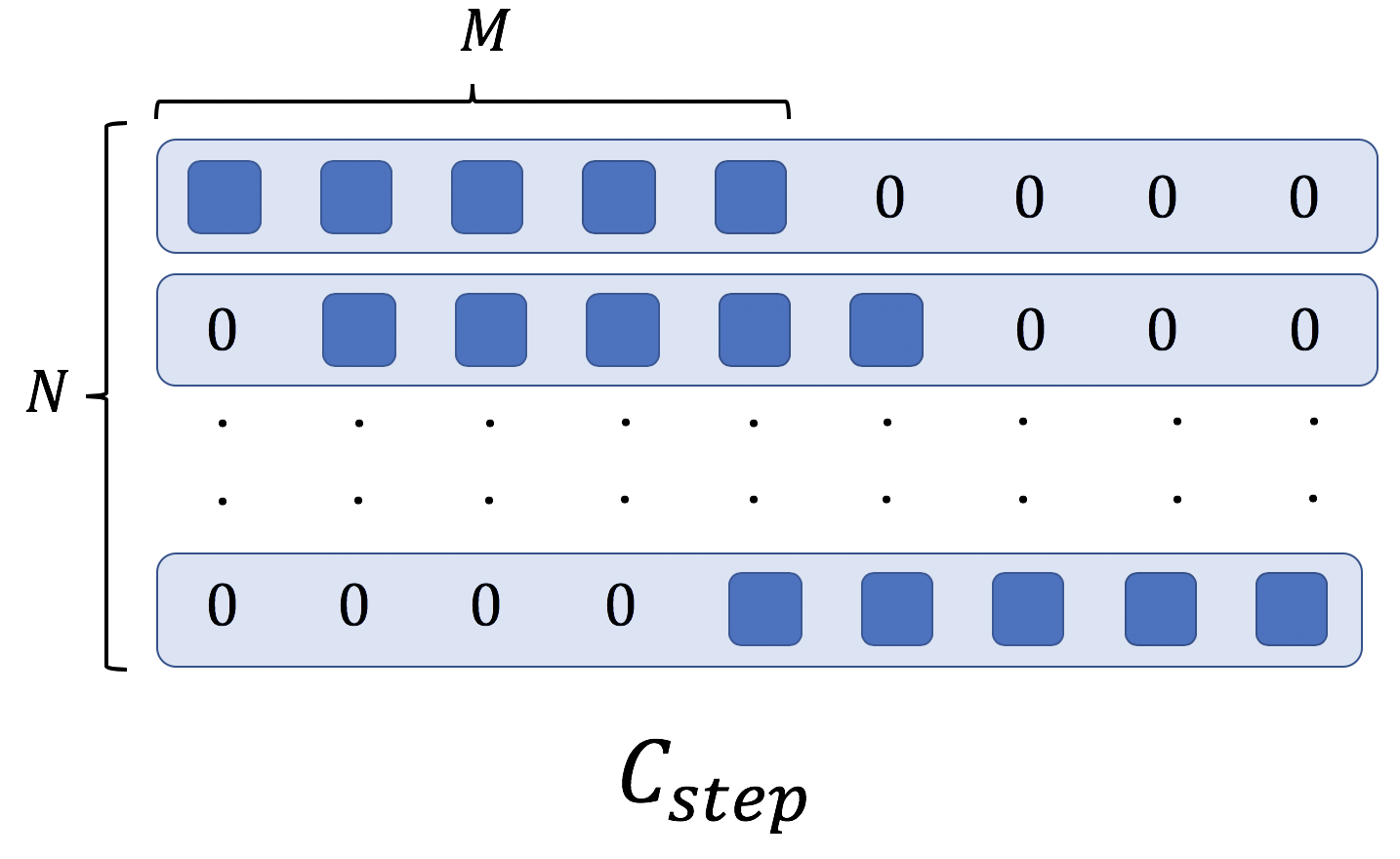}
\caption{Stepped packed weight matrix $\mathbf{C}'$. With this preprocessing on plaintext, no rotation required on ciphertext.}
\label{fig:ourtransitionmethod3}
\end{centering}
\end{figure}

In addition, if we prepare the packed matrix with pre-rotated ``stepped'' packing, which means that we prepare $\mathbf{C}'=\texttt{rot}\big(C_i,i\big)$ as preprocessed weight matrix before encryption as shown in figure~\ref{fig:ourtransitionmethod3}. This preprocess can result in the reduction of on the fly rotation operation over encrypted weight. Algorithm \ref{transpose-diagonal-with-preprocess-rot} shows the procedure of this method.

which has total rotation reduced down to $(M+1)$ .
Here, table~\ref{table:outtransitionmethod1} summarizes the computation cost for each method.

\begin{longtable}{|l||c|c|}
\hline
\textbf{Packing Method}    & \textbf{Multiplication} & \textbf{Rotation} \\ \hline
\endfirsthead
\textbf{Packing Method}    & \textbf{Multiplication} & \textbf{Rotation} \\ \hline
\endhead
\multicolumn{3}{r}{Continued on next page}
\endfoot
\endlastfoot
Row Packing                & $N$                       & $NM$                \\
Diagonal Packing           & $0$                       & $qM$                \\
Diagonal Packing with Step & $0$                       & $M$                 \\ \hline
\caption{Computation cost for taking matrix transpose. Multiplication and rotation for ciphertext is hugely reduced with our algorithm.}\label{table:outtransitionmethod1}
\end{longtable}

 \begin{breakablealgorithm}\label{transpose-diagonal}
 \caption{Algorithm of taking transpose with diagonal packing}
 \begin{algorithmic}[1]
 \raggedright
 \Require $\textbf{C}=\{\textbf{c}_i,0\le i\le N-1 \}$
 \Ensure  $\textbf{D}=\{\textbf{d}_i,0\le i\le M-1 \}$
 \State {$e_i =\left\{
\begin{array}{ll}
q& \text{if $0\leq i \leq M-r$} \\
q+1 &\text{if $M-r < i \leq M$}
\end{array}\right.$}
  \State $\textbf{d}_i = \sum_{s=0}^{s=e_i}{\texttt{rot}\big(C_{sM-i}, sM-i\big)}\quad\text{where $N = qM + r$.}$
 \State{\textbf{RETURN} $\textbf{D}$} 
 \end{algorithmic} 
 \end{breakablealgorithm}
 
 \begin{breakablealgorithm}\label{transpose-diagonal-with-preprocess-rot}
 \caption{Algorithm of taking transpose with diagonal packing}
 \begin{algorithmic}[1]
 \raggedright
 \Require $\textbf{C'}=\texttt{rot} (C, i)\quad \text {for} \quad 0 \le i \le n, \quad \text{where} \quad \textbf{C}=\{\textbf{c}_i,0\le i\le N-1 \}$
 \Ensure  $\textbf{D}=\{\textbf{d}_i,0\le i\le M-1 \}$
 \State {$e_i =\left\{
\begin{array}{ll}
q& \text{if $0\leq i \leq M-r$} \\
q+1 &\text{if $M-r < i \leq M$}
\end{array}\right.$}
  \State $\textbf{d}_i = \texttt{rot}\big(C_{N-i}, -i\big) +  \sum_{s=1}^{s=e_i}{C_{sM-i} } \quad\text{where $N = qM + r$.}$
 \State{\textbf{RETURN} $\textbf{D}$} 
 \end{algorithmic} 
 \end{breakablealgorithm}
 
Once this algorithm is applied, packed matrix of weights is used to feedforward, and is then transposed to propagate backward to update weights. As shown in~\cite{KimSonKimLeeChe_2018}, generated updates of the weights in this transition method has the same structure as the original weight structure. Therefore, update of the weights can also be done in server side. The advantage of this is that we can apply bootstrapping ~\cite{chen2019improved} in order to complete the training process without sending ciphertext back to the client. However, in our implementation, we send the ciphertext back to client for weight update for each batch iteration for simplicity. It is important here to understand very well the underlying computational tasks in order to fit a good cryptographic solution. Here, to show the concrete example, we consider the example of a neural network with input size $6$, hidden size $3$, and output size $1$. Table~\ref{table:ourtransitionmethod2} shows the number of ciphertext multiplications and rotations needed during the training procedure: (1) by naively using row packing and (2) by using diagonal packing with our transpose method. We observe that the total number of rotations and multiplications is reduced using our method by almost $3$ folds with multiplication circuit depth kept minimum.

\newpage

\begin{longtable}{|llll|ll|ll|}
\cline{5-8}
\multicolumn{4}{c|}{} & \multicolumn{2}{c|}{Diagonal} & \multicolumn{2}{c|}{Row}  \\
\cline{5-8}
\multicolumn{4}{c|}{}   & \texttt{mult} & \texttt{rot} &       \texttt{mult} & \texttt{rot}      \\ \hline
\endfirsthead
\cline{5-8}
\multicolumn{4}{c|}{} & \multicolumn{2}{c|}{Diagonal} & \multicolumn{2}{c|}{Row}  \\
\cline{5-8}
\multicolumn{4}{c|}{}  & \texttt{mult} & \texttt{rot} &       \texttt{mult} & \texttt{rot}     \\ \hline
\endhead
\multicolumn{8}{r}{\graytext{Continued on next page}}
\endfoot
\endlastfoot
FF         & Dense       & $6$ & $3$ & $6$    & $3$   &       $3$    & $18$       \\
           & square actv &   &   & $1$    & $0$   &       $1$    & $0$        \\
           & Dense       & $3$ & $1$ & $3$    & $1$   &       $1$    & $3$        \\
           & square actv &   &   & $1$    & $0$   &       $1$    & $0$        \\ \hline
Transition & transition  & $6$ & $3$ & $0$    & $3$   &       $18$   & $18$       \\
           &             & $3$ & $1$ & $0$    & $1$   &       $3$    & $3$        \\ \hline
BP         & Dense       & $1$ & $3$ & $1$    & $3$   &       $3$    & $3$        \\
           & square actv &   &   & $1$    &     &       $1$    & $0$        \\
           & Dense       & $3$ & $6$ & $3$    & $6$   &       $6$   & $18$       \\
           & square actv &   &   & $1$    & $0$   &       $1$    & $0$        \\ \hline
Total      &             &   &   & $17$   & $17$  &       $38$   & $63$   \\ \hline
\caption{Example of number of necessary rotations and multiplications with the simple neural network model. First layer has $6$ inputs. Hidden layer has dimension $3$. Output layer has dimension $1$. Total computation cost reduced from $101$ to $34$.}
\label{table:ourtransitionmethod2}
\end{longtable}

As can be seen, total number of multiplication and rotation reduced from $101$ to $34$ using our method with diagonal packing and huge reduction comes from the optimization of transition process. These reduction can be more significant with larger matrix size, which is the usual case for real world application.

\section{Implementation}\label{sec:Sec5}

To verify the advantage of our method, we implemented a neural network model with one hidden layer for the iris dataset classification. Iris dataset is a well-known tutorial dataset for machine learning, which contains $150$ datasets and $150$ labels of $3$ kinds of iris in total. Each element of the dataset consists four iris properties that are flagged for membership. We design our neural network to have one hidden layer as in~\cite{CryptoNets_2016, KimSonKimLeeChe_2018} and described as in table~\ref{table:impl1}.

\begin{longtable}{|l|l|l||l|l|}
\hline
\textbf{layer name} & \textbf{dim input} & \textbf{dim output} &  \textbf{dataset}  & iris \\ \hline
\endfirsthead
\hline
\textbf{layer name} & \textbf{dim input} & \textbf{dim output} &  \textbf{dataset}  & iris \\ \hline
\endhead
\endlastfoot
\hline
\multicolumn{5}{|r|}{\graytext{Continued on the following page.}}\\ \hline
\endfoot
Dense               & $4$                   & $10$                   &  Learning Rate & $0.1$  \\
Square Activation   & $10$                  & $10$                   &  Batch Size    & $1, 20$ \\
Dense               & $10$                & $3$                    &  Loss Function & SGD \\
Square Activation   & $3$                   & $3$                    &  Epochs        & $400$ \\ \hline
\caption{Our method on the iris dataset. Square activation is friendly with respect to homomorphic operations in order to keep the multiplication depth as low as possible. A batch size $1$ is used for performance comparison between row method and our diagonal method. (proportion train:test $= 0.8:0.2$)}
\label{table:impl1}
\end{longtable}

\begin{figure}[h]
\begin{centering}
\includegraphics[width=0.75\textwidth]{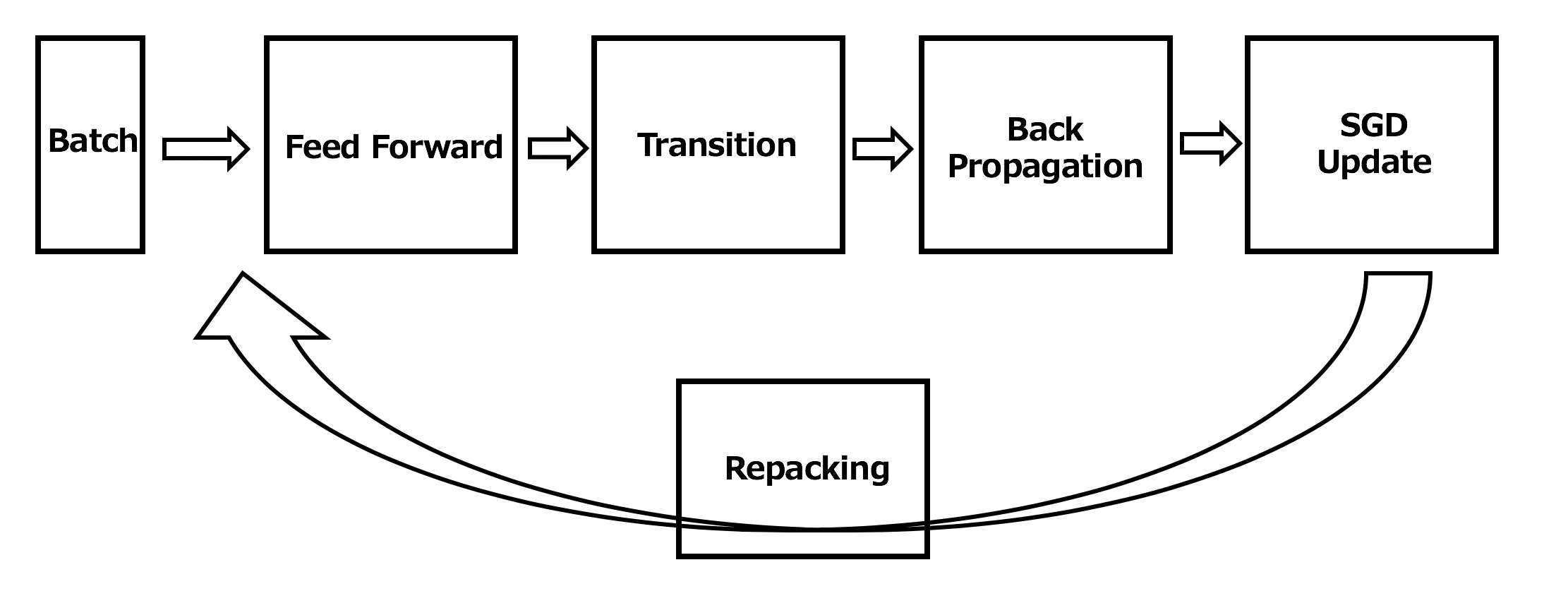}
\caption{Training flow per batch.}
\label{fig:impl1}
\end{centering}	
\end{figure}

Figure~\ref{fig:impl1} shows the execution flow of training. Repacking of the weight matrix is done in order to prevent the growth of noise per batch training.
The repacking is done by transferring encrypted data to a trusted environment temporarily to perform decryption and re-encrypt of the weight matrix that is sent back to the server. However, as we mentioned earlier, this communication is not necessary when bootstrapping is applied.

We use Microsoft SEAL library as a back end. Encryption parameters for the degree of modulus is $2^{15}$ and bit sequence of modulus chain is to be in the multiset \mbox{$\{60, 40, 40, 40, 40, 40, 40, 40, 40, 40, 60\}$} ($9$ levels of modulus chains available. On the other hand, encryption parameters for the row method has $12$ levels, because of the extra multiplication with the unit vector. The machine specifications for this experiment are: Intel(R) Core(TM) i7-8700K CPU @ $3.70$GHz, $32$GB RAM and $6$ cores. 

Table~\ref{table:impl2} summarizes the running time for each steps of training: (1) feedforward, (2) taking transpose, (3) backpropagation, (4) SGD update and (5) row repacking versus diagonal repacking. For simplicity, the batch size is set to one for this comparison. We observe that the entire process gets faster by $3$ times with our optimization.

\begin{table}[h]
\begin{centering}
\begin{tabular}{|l|l|l|}
\hline
\textbf{process} & \textbf{row} &   \textbf{diagonal}      \\ \hline
Feedforward      &      $12.42$  &       $3.10$        \\
Transition       &     $7.28$    &        $2.15$    \\
Backpropagation  &      $5.46$    &      $3.12$         \\
SGD Update       &      $0.23$     &       $0.23$         \\
Repacking       &       $0.62$     &        $0.50$            \\ \hline
Total           &   $28.47$      &       $9.25$ \\ \hline
\end{tabular}
\caption{Computation time for each section of training per one iteration with batch size being 1. Time took for row packing method and diagonal packing method is shown side by side. Time unit in second.}
\label{table:impl2}
\end{centering}
\end{table}

\begin{figure}[h]
\begin{centering}
    \includegraphics[width=1.0\textwidth]{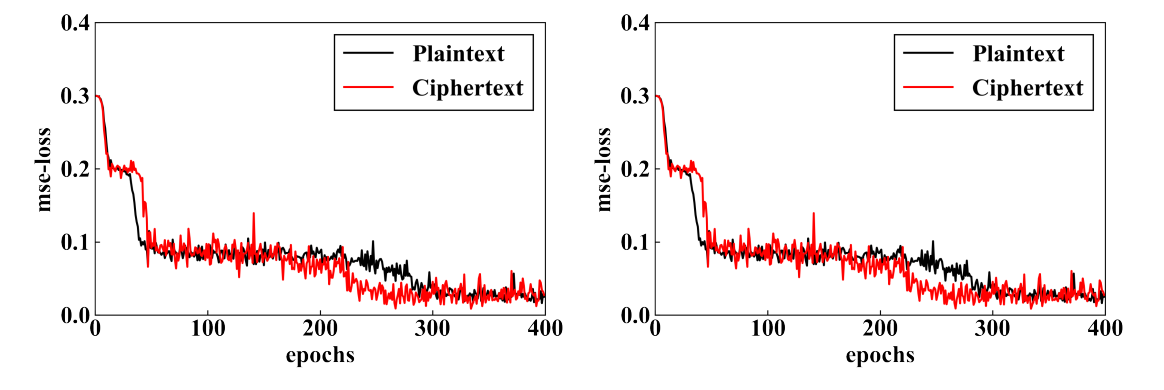}
    \caption{Mean squared error loss of training dataset (left) and test dataset (right) with comparison between training with CKKS ciphertext (red) and plaintext(black)}
    \label{fig:impl2}
\end{centering}
\end{figure}

\begin{figure}[h]
\begin{centering}
	\includegraphics[width=1.0\textwidth]{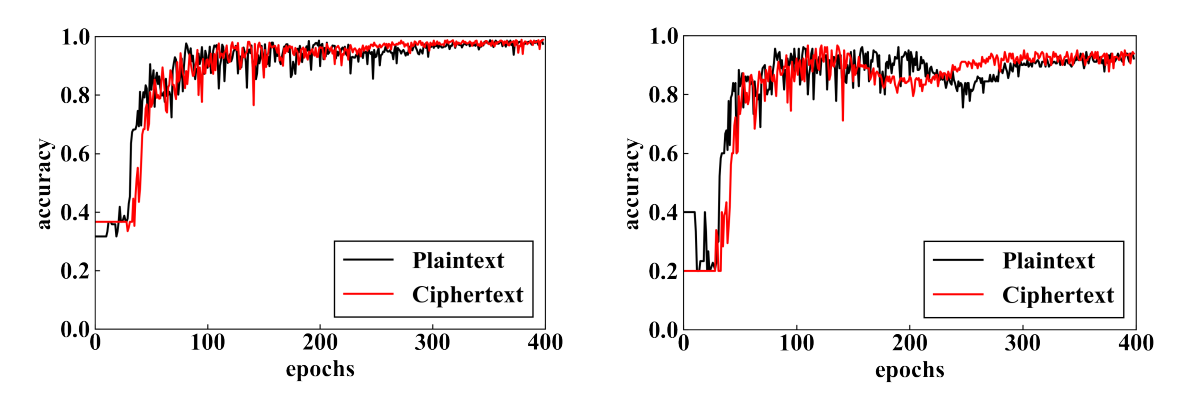}
	\caption{Accuracy of training dataset (left) and that of test dataset (right) with comparison between training with CKKS ciphertext(red) and plaintext (black).}
	\label{fig:impl3}
\end{centering}		
\end{figure}

To assert the correctness of our method, training is done entirely with our method. We use same model architecture and a batch size of $20$ instead. Batch process is parallelized by Open MP with $12$ threads, which gave us around twice computation performance. Total time for entire process was $107,520$ seconds which is about $29.8$ hours with $400$ epochs. In addition, we wrote a program of training this model with plaintext using same architecture. This code is used to compare the accuracy of a model trained over plaintext and a model trained over ciphertext through CKKS. The mean square loss function and the learning accuracy transition curve per epoch are plotted side by side on figures~\ref{fig:impl2} and~\ref{fig:impl3}. After $400$ epochs, final loss function is $0.0249$ for plaintext-based model and $0.273$ for ciphertext-based model. On the other hand, final accuracy is $0.9805$ for plaintext-based model and $0.9847$ for ciphertext-based model. As can be seen, intrinsic approximation of real number of CKKS scheme did not affect the model accuracy.

\section{Conclusion and further research}\label{sec:Sec6}

The needs for privacy-preserving solutions in untrusted environments in which machine learning algorithms are executed are essential nowadays. To achieve such privacy-preserving algorithms, homomorphic encryption has gone out of academia to be now operational at an industrial scale. We proposed an efficient privacy-preserving neural network training architecture that relies on a packing method to optimize the transposition of matrices. As a consequence, training neural networks homomorphically yields to an improvement about $3$ times with respect to other architectures. We tested its advantages on the well-known iris dataset using the SEAL library.
We are investigating how our method can be used to train more complex neural network. Homomorphic encryption based neural network training is still a big challenge in terms of computation, however, we believe that it can be a huge breakthrough for privacy preserving technology in coming years.


\bibliographystyle{plain}

\bibliography{train_nn}

\end{document}